\documentclass[conference]{IEEEtran}
\IEEEoverridecommandlockouts

\usepackage{cite}
\usepackage{amsmath,amssymb,amsfonts}
\usepackage{algorithmic}
\usepackage{graphicx}
\usepackage{textcomp}
\usepackage{xcolor}
\usepackage{subfig}
\def\BibTeX{{\rm B\kern-.05em{\sc i\kern-.025em b}\kern-.08em
    T\kern-.1667em\lower.7ex\hbox{E}\kern-.125emX}}
\begin{document}

\title{\huge A Comparison among Single Carrier, OFDM, and OTFS in mmWave Multi-Connectivity Downlink Transmissions\\
}
\author{
	\IEEEauthorblockN{Fabian G\"{o}ttsch\IEEEauthorrefmark{1}\IEEEauthorrefmark{2}, 
     Shuangyang Li\IEEEauthorrefmark{1},
     Lorenzo Miretti\IEEEauthorrefmark{1}\IEEEauthorrefmark{3},
            Giuseppe Caire\IEEEauthorrefmark{1}\IEEEauthorrefmark{2} and S{\l}awomir Sta{\'n}czak\IEEEauthorrefmark{1}\IEEEauthorrefmark{3}
	}
\IEEEauthorblockA{\IEEEauthorrefmark{1}Faculty of Electrical Engineering and Computer Science, Technical University of Berlin, Berlin, Germany}
\IEEEauthorblockA{\IEEEauthorrefmark{2}Massive Beams, Berlin, Germany}
\IEEEauthorblockA{\IEEEauthorrefmark{3}Fraunhofer Institute for Telecommunications Heinrich-Hertz-Institut, Berlin, Germany}
\IEEEauthorblockA{E-mail: \{fabian.goettsch, shuangyang.li, miretti, caire\}@tu-berlin.de, slawomir.stanczak@hhi.fraunhofer.de}

 \vspace{-.07cm}
}
  
\maketitle

\begin{abstract}
In this paper, we perform a comparative study of common wireless communication waveforms, namely the single carrier (SC), orthogonal frequency-division multiplexing (OFDM), and orthogonal time-frequency-space (OTFS) modulation in a millimeter wave (mmWave) downlink multi-connectivity scenario, where multiple access points (APs) jointly serve a given user under imperfect time and frequency synchronization errors. For a fair comparison, all the three waveforms are evaluated using variants of common frequency domain equalization (FDE). To this end, a novel cross domain iterative detection for OTFS is proposed. The performance of the different waveforms is evaluated numerically in terms of pragmatic capacity. The numerical results show that OTFS significantly outperforms SC and OFDM at cost of reasonably increased complexity, because of the low cyclic-prefix (CP) overhead and the effectiveness of the proposed detection.
\end{abstract}


\section{Introduction}
Distributed multiple-input multiple-output (MIMO) systems, such as cell-free massive MIMO, have been identified as key enabling technologies for supporting ultra high data rates required by next generation wireless networks~\cite{ngo2024ultradense}. The idea of distributed MIMO is to combine the advantages of multi-user massive MIMO systems \cite{Caire-Shamai-TIT03, Marzetta-TWC10} with the benefits of simple cooperative multi-connectivity techniques and ultra-dense network deployments, thus enabling unprecedented spectral efficiency and macro-diversity gains (see \cite{ngo2024ultradense} and references therein).
The macro-diversity gains are particularly important in mmWave and sub-THz networks, where consistently high data rates are achievable in practice if the user equipment (UE) is connected to at least one access point (AP) via a reflected or line-of-sight (LoS) path with large channel gain \cite{miretti2024robust}. 

While the benefits of multi-connectivity and distributed MIMO systems compared to conventional collocated MIMO systems are clear, deploying and operating a network with multiple cooperating APs is also known to present several challenges.
Critical issues in distributed MIMO systems with multiple APs jointly serving a given UE are the carrier frequency offset (CFO) caused by imperfect oscillator synchronization, and time offset (TO) essentially caused by imperfect compensation of the propagation delays \cite{irmer2011coordinated}. In general, CFO and TO occur in any wireless communication system, collocated or distributed. 
However, these effects are much more pronounced when a UE receives signals from different APs. This is because each link has a different frequency mismatch between the (multiple) transmit and receive oscillators. 
In addition, different geographical locations of the APs may lead to significantly different delay and Doppler profiles at the UE side, i.e., each AP-UE link is characterized by a different delay-Doppler (DD) pair. 
Note that both Doppler and CFO introduce time-varying phase shifts to the received signal, while both delay and TO cause time shifts to the received signal.
Even with some sort of DD shift pre-compensation at each AP, such as~\cite{you2020network}, perfect pre-compensation is very difficult to achieve in practice and the UE still suffers from distortions due to the residual phase and time shifts. 

In this paper, we are interested in evaluating different waveforms in the multi-connectivity scenario described above, where the received signal at the UE is distorted by small phase and time shifts after pre-compensation. 
Our evaluation will include three widely studied waveforms, namely, the single carrier (SC), the orthogonal frequency-division multiplexing (OFDM), and the orthogonal time frequency space (OTFS) waveform. The consideration of the OTFS waveform is motivated by its known robustness against channel distortions caused by delay and Doppler thanks to the delay-Doppler (DD) domain symbol placement~\cite{Li2020performance}.
To the best knowledge of the authors, no thorough comparison of such is presented in the literature. Related studies compare OFDM and OTFS in a cell-free \cite{mohammadi2022cell} and non-terrestrial \cite{buzzi2023leo} network with multiple APs serving a given UE. However, both works consider sub-6 GHz communication and do not include SC modulation in their comparison, which is a suitable choice for high carrier frequencies \cite{miretti2024little}.

This paper carries out such a comparison in a fair way by restricting the analysis to reduced-complexity detection methods for the three waveforms based on variations of the single-tap frequency domain equalization (FDE) method, which is known to yield promising performance in the presence of small channel Doppler. Particularly, we propose a novel cross-domain iterative detection (CDID) for OTFS based on the FDE, where extrinsic information of the information symbols is passed via the unitary transform. In contrast to existing CDID methods \cite{li2021cross,Mengmeng2023crossdomain}, the proposed scheme calculates the extrinsic information from the time domain instead of the frequency domain directly after the FDE. Such an operation can significantly improve the quality of the extrinsic information and it is particularly suitable for scenarios with a diagonal-dominant \textit{frequency} domain channel matrix\footnote{Note that here we refer to the frequency domain channel matrix accounting for inter-carrier and/or inter-symbol interference, not to a diagonal matrix characterizing a MIMO channel.}. Our evaluation shows that OTFS outperforms both SC and OFDM in terms of pragmatic capacity at the cost of a relatively small increase in detection complexity due to the cross domain iteration. 

\emph{Notations:} The blackboard bold letters ${\mathbb{E}}$ and ${\mathbb{C}}$ denote the expectation operator and complex number field, respectively; the notations $(\cdot)^{\rm{T}}$ and $(\cdot)^{\rm{H}}$ denote the transpose and the Hermitian transpose for a matrix, respectively; 
$\textrm{vec}(\cdot)$ denotes the vectorization operation; 
${{{\bf{F}}_N}}$ and ${{{\bf{I}}_M}}$ denote the discrete Fourier transform (DFT) matrix of size $N\times N$ and the identity matrix of size $M\times M$; $x[l]$ denotes the $l$-th entry of the vector $\bf x$ and $a_{i,j}$ denotes the $(i,j)$-th element of the matrix $\bf A$.

\section{System Models for Downlink Transmission with Multi-Connectivity}
\begin{figure}[t]
\centering
\includegraphics[width=0.8\linewidth]{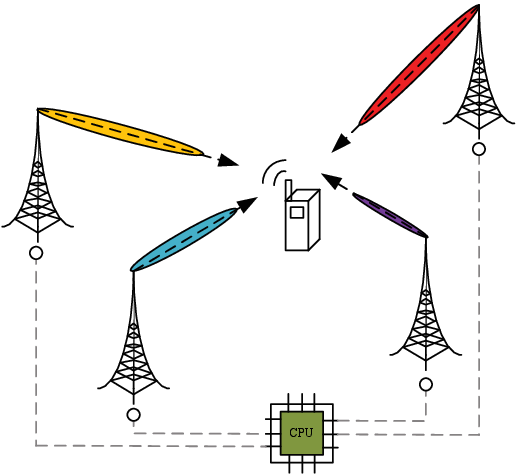}
\caption{Considered system model for downlink transmission with multi-connectivity.}
\label{system_model}
\centering
\vspace{-2mm}
\end{figure}
We consider a downlink multi-connectivity scenario with $M_{\rm AP}$ APs jointly serving a single user in the same time-frequency resource as shown in Fig.~\ref{system_model}. We assume that each AP 
is capable of steering a very narrow beam towards the UE, while the UE is equipped with an omnidirectional antenna. 
To model the crucial impact of LoS signal blockage at mmWave and sub-THz frequencies, we characterize the connectivity between the UE and the $i$-th AP via a Bernoulli distributed random variable $h_i$, $1 \le i \le M_{\rm AP} $ with a blocking rate $q$~\cite{shokri2015millimeter, miretti2024robust}, i.e., with probability $q$ being zero and probability $1-q$ being one.
Due to the presence of TO and CFO, the same information signal $s(t)$ will be superimposed non-coherently at the UE side, yielding the received signal $r(t)$ of the form
\begin{align}
r\left( t \right) = \sum\nolimits_{i = 1}^{{M_{{\rm{AP}}}}} {{h_i}{e^{j2\pi {\nu _i}\left( {t - {\tau _i}} \right)}}s\left( {t - {\tau _i}} \right)} + w\left( t\right) ,
\label{io_time_cont}
\end{align}
where $w\left( t\right)$ is the additive white Gaussian noise (AWGN) process with one-sided power spectral density (PSD) $N_0$, and ${\tau _i}$ and ${\nu _i}$ are the effective TO and effective CFO incorporating the channel Doppler after pre-compensation associated to the link between the $i$-th AP and the user, respectively. Here, both TO and effective CFO are assumed to be \textit{small} compared to the symbol duration and subcarrier spacing after pre-compensation at each AP. In addition, note that, for the sake of analysis, the impact of multi-path fading is neglected in \eqref{io_time_cont}. This is a reasonable approximation under the considered highly directional LoS beamformed transmission in the noise limited regime, which is the typical scenario in mmWave/subTHz mobile access systems \cite{miretti2024little}.

\begin{figure}[t]
\centering
\includegraphics[width=0.8\linewidth]{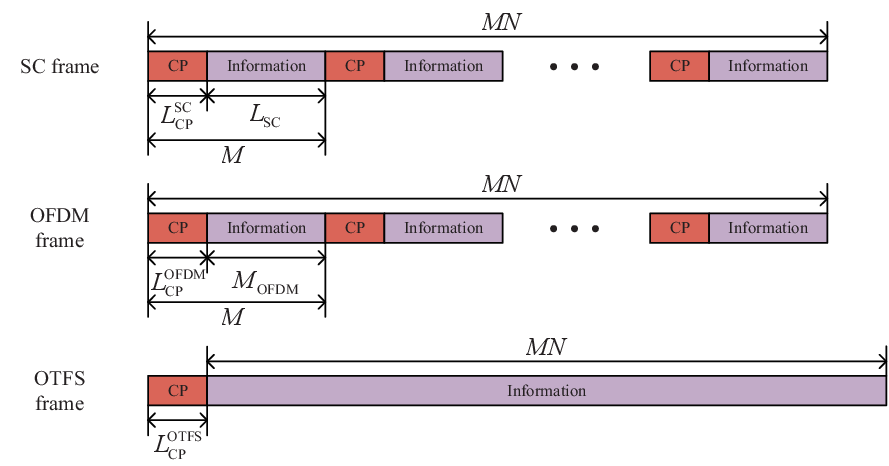}
\caption{Considered frame structures for SC, OFDM, and OTFS. }
\label{Frame_structures}
\centering
\vspace{-2mm}
\end{figure}

We consider block-wise time domain frame structures for both SC and OFDM as shown in Fig.~\ref{Frame_structures}, where adjacent data blocks are separated by a cyclic prefix (CP) of lengths $L_{\rm CP}^{\rm SC}$ and $L_{\rm CP}^{\rm OFDM}$, respectively. The adopted CP allows both SC and OFDM frames to have multiple independent blocks, and the received signals after CP removal to be characterized by an approximately circulant convolution matrix.

Without loss of generality, we assume that both SC and OFDM use $N$ blocks per frame and each block contains $M$ symbols.
On the other hand, we apply the commonly adopted reduced-CP structure~\cite{Raviteja2019practical} to OTFS as shown in Fig.~\ref{Frame_structures}, where the CP length is given by $L_{\rm CP}^{\rm OTFS}$. The detailed parameters for the considered three frame structures are summarized in Table~\ref{frame_parameters}. As shown in Table~\ref{frame_parameters}, OTFS frame is slightly larger than the other two but it occupies a much smaller CP overhead. 
In what follows, we will briefly introduce the transceivers for the considered three waveforms as depicted in Fig.~\ref{transceiver_diagrams}.

\begin{table}[]
\vspace{0.03in}
\caption{Parameters of different frame structures. }
\centering
\begin{tabular}{|c|c|c|c|}
\hline
     & \begin{tabular}[c]{@{}c@{}}\textbf{SC}\end{tabular} & \begin{tabular}[c]{@{}c@{}}\textbf{OFDM}\end{tabular} & \begin{tabular}[c]{@{}c@{}}\textbf{OTFS}\end{tabular}  \\ \hline
Single CP length & $L_{\rm CP}^{\rm SC}$ & $L_{\rm CP}^{\rm OFDM}$ & $L_{\rm CP}^{\rm OTFS}$ \\ \hline
Inf. symbols per block & $L_{\rm SC}$ & $M_{\rm OFDM}$ & $MN$ \\ \hline
Number of blocks & $N$ & $N$ & $1$ \\ \hline
Total CP length & $L_{\rm CP}^{\rm SC}N$ & $L_{\rm CP}^{\rm OFDM}N$ & $L_{\rm CP}^{\rm OTFS}$ \\ \hline
\begin{tabular}[c]{@{}c@{}}Number of\\transmitted symbols\end{tabular} & $MN$ & $MN$ & $L_{\rm CP}^{\rm OTFS}+MN$ \\ \hline
\end{tabular}
\label{frame_parameters}
\end{table}

\subsection{Single Carrier}
As shown in Fig.~\ref{transceiver_diagrams}, a length-$L_{\rm SC}N$ information symbol vector ${{\bf{x}}_{{\rm{SC}}}}$ is transmitted using the SC waveform, where $L_{\rm SC} = M - L_{\rm CP}^{\rm SC}$. After appending CPs according to Fig.~\ref{Frame_structures}, the resultant vector ${{\bf{\tilde x}}_{{\rm{SC}}}}$ of length $MN$ is passed to the transmit filter. 
Let $p\left( t \right)$ be the transmit shaping pulse, which is $T_s$-orthogonal with $T_s$ being the symbol period. Then the transmitted signal with the SC waveform can be written as
{\color{black}${s_{{\rm{SC}}}}\left( t \right) =\sum\nolimits_{n = 0}^{MN - 1} {{{\tilde x}_{{\rm{SC}}}}\left[ n \right]p\left( {t - n{T_s}} \right)}$.}
Using~\eqref{io_time_cont}, the received signal is then given by 
\begin{align}
{r_{{\rm{SC}}}}\left( t \right) &= \sum\limits_{i = 1}^{{M_{{\rm{AP}}}}} {{h_i}{e^{j2\pi {\nu _i}\left( {t - {\tau _i}} \right)}}{s_{{\rm{SC}}}}\left( {t - {\tau _i}} \right)}  + w\left( t\right)\notag\\
& \hspace{-.95cm} = \sum\limits_{i = 1}^{{M_{{\rm{AP}}}}} {\sum\limits_{n = 0}^{MN - 1} {{h_i}{e^{j2\pi {\nu _i}\left( {t - {\tau _i}} \right)}}{{\tilde x}_{{\rm{SC}}}}\left[ n \right]p\left( {t \!-\! n{T_s}\! -\! {\tau _i}} \right)} }\!+ \!w\left( t\right).
\label{SC_receive_signal}
\end{align}
By passing the received signal to the matched filter ${p^*}\left( t \right)$, a set of sufficient statistics 
is obtained for detection. Let us define
the ambiguity function of ${p}\left( t \right)$ with respect to delay $\tau$ and Doppler $\nu$ as ${A_p}\left( {\tau ,\nu } \right) \buildrel \Delta \over = \int_{ - \infty }^\infty  {p\left( t \right)} {p^*}\left( {t - \tau } \right){e^{ - j2\pi \nu \left( {t - \tau } \right)}}{\rm{d}}t$
and the effective channel coefficient characterizing the correlation between the $n$-th transmit symbol ${{{\tilde x}_{{\rm{SC}}}}\left[ n \right]}$ and the $m$-th receive symbol ${{{\tilde y}_{{\rm{SC}}}}\left[ m \right]}$ with respect to the $i$-th AP as
\begin{align}
g_{m,n}^{\left( i \right)} \triangleq {h_i}{e^{j2\pi n{\nu _i}{T_s}}}A_p^*\left( {\left( {n - m} \right){T_s} + {\tau _i},{\nu _i}} \right) ,
\label{effective_channel_coef}
\end{align}
 where, denoting $w\left[ m \right]$ as the $m$-th AWGN sample,
\begin{align}
{{\tilde y}_{{\rm{SC}}}}\left[ m \right] &= \int_{ - \infty }^\infty  {{r_{{\rm{SC}}}}\left( t \right)} {p^*}\left( {t - m{T_s}} \right){\rm{d}}t \notag\\
&=  \sum\nolimits_{n = 0}^{MN - 1} {{{\tilde x}_{{\rm{SC}}}}\left[ n \right]\sum\nolimits_{i = 1}^{{M_{{\rm{AP}}}}} {g_{m,n}^{\left( i \right)}} } + w\left[ m \right]
\label{SC_io_withCP}.
\end{align}
We can rewrite \eqref{SC_io_withCP} in matrix form as
{\color{black}${{{\bf{\tilde y}}}_{{\rm{SC}}}} = \sum\nolimits_{i = 1}^{{M_{{\rm{AP}}}}} {{{\bf{G}}^{\left( i \right)}}} {{{\bf{\tilde x}}}_{{\rm{SC}}}}+{\bf w}$,}
where ${{{\bf{\tilde y}}}_{{\rm{SC}}}} = \left[ {\tilde y}_{\rm SC}[0], \dots, {\tilde y}_{\rm SC}[MN-1] \right]$, ${{{\bf{w}}}} = \left[ w[0], \dots, w[MN-1] \right]$ and ${{{\bf{G}}^{\left( i \right)}}}$ is the time domain effective channel corresponding to the transmission of the $i$-th AP, whose $\left( {m,n} \right)$-th element is ${g_{m,n}^{\left( i \right)}}$.
Let ${\bf{A}}_{{\rm{CP}}}^{{\rm{SC}}}$ and ${\bf{R}}_{{\rm{CP}}}^{{\rm{SC}}}$ be the SC CP addition and removal matrices of dimensions $MN \times L_{\rm SC}N$ and $L_{\rm SC}N \times MN$, respectively, whose structures will be discussed in details in the following section.
Then, we arrive at the time-domain input-output relation between the information symbols ${{{\bf x}_{{\rm{SC}}}}}$ and the received symbols after CP removal ${{{\bf y}_{{\rm{SC}}}}}$ given by 
\begin{align}
{{\bf{y}}_{{\rm{SC}}}} = \sum\nolimits_{i = 1}^{{M_{{\rm{AP}}}}} {{\bf{R}}_{{\rm{CP}}}^{{\rm{SC}}}{{\bf{G}}^{\left( i \right)}}} {\bf{A}}_{{\rm{CP}}}^{{\rm{SC}}}{{\bf{x}}_{{\rm{SC}}}} + {\bf{w}}.
\label{SC_io_withoutCP_mtx}
\end{align}

\subsection{OFDM}
\begin{figure}[t]
\centering
\includegraphics[width=0.9\linewidth]{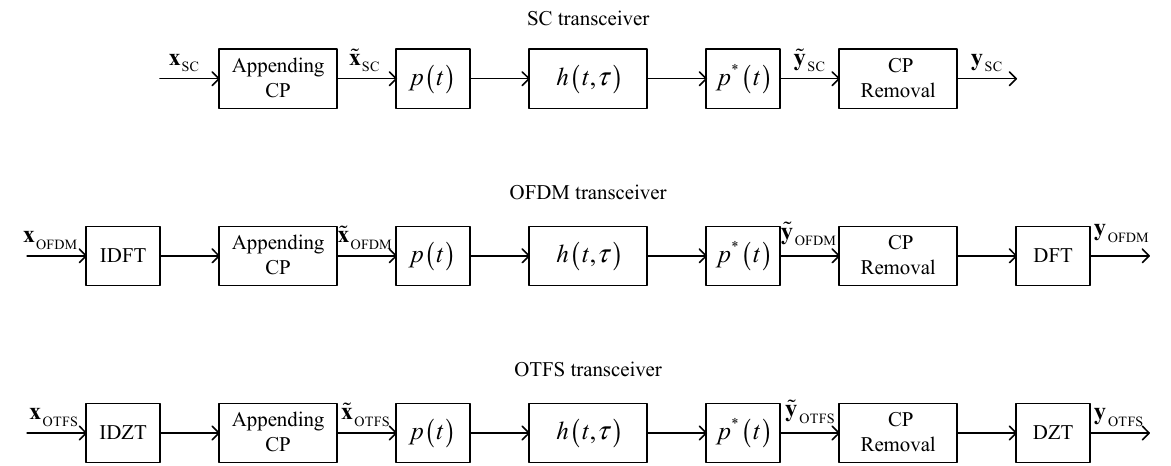}
\caption{Transceiver diagrams for SC, OFDM, and OTFS. }
\label{transceiver_diagrams}
\centering
\vspace{-5mm}
\end{figure}

Let ${\bf X}_{\rm OFDM}$ of size $M_{\rm OFDM} \times N$ be the information symbol matrix for OFDM, where $M_{\rm OFDM}$ denotes the number of sub-carriers and $N$ is the number of OFDM symbols which is equal to the number of blocks for the SC waveform. Then, the time domain OFDM symbol vector of length $MN$ can be obtained according to Fig.~\ref{transceiver_diagrams}, which can be written by
{\color{black}${{{\bf{\tilde x}}}_{{\rm{OFDM}}}} = {\bf{A}}_{{\rm{CP}}}^{{\rm{OFDM}}}{\left({{\bf{I}}_{{N}}} \otimes {\bf{F}}_{{M_{{\rm{OFDM}}}}}^{\rm{H}}\right)}{{\bf{x}}_{{\rm{OFDM}}}}$,}
where ${{\bf{x}}_{{\rm{OFDM}}}} = {\rm{vec}}\left( {{{\bf{X}}_{{\rm{OFDM}}}}} \right)$ is the vectorized version of ${{{\bf{X}}_{{\rm{OFDM}}}}}$ and ${\bf{A}}_{{\rm{CP}}}^{{\rm{OFDM}}} \in \mathbb{R}^{MN \times M_{\rm OFDM}N}$ is the CP addition matrix for OFDM that will be explained later.
Let ${\bf{C}}_{{\rm{CP}}}^{{\rm{OFDM}}} \in \mathbb{R}^{M_{\rm OFDM}N \times MN}$ be the CP removal matrix for OFDM. After some straightforward derivations, we arrive at the frequency domain input-output relation
\begin{align}
{{\bf{y}}_{{\rm{OFDM}}}} = &\sum\nolimits_{i = 1}^{{M_{{\rm{AP}}}}} {{\left({{\bf{I}}_{{N}}} \otimes {\bf{F}}_{{M_{{\rm{OFDM}}}}}\right)} {\bf{R}}_{{\rm{CP}}}^{{\rm{OFDM}}}{{\bf{G}}^{\left( i \right)}}}\notag\\
&{\bf{A}}_{{\rm{CP}}}^{{\rm{OFDM}}} {\left({{\bf{I}}_{{N}}} \otimes {\bf{F}}_{{M_{{\rm{OFDM}}}}}^{\rm{H}}\right)}{{\bf{x}}_{{\rm{OFDM}}}} + {\bf{w}}
\label{OFDM_io_withoutCP_mtx}.
\end{align}
In~\eqref{OFDM_io_withoutCP_mtx}, we use the same vector $\bf w$ representing the AWGN samples as in~\eqref{SC_io_withoutCP_mtx} since they share the same distribution.

\subsection{OTFS}
We consider the inverse discrete Zak transform (IDZT)-based OTFS implementation~\cite{lampel2021orthogonal} as shown in Fig.~\ref{transceiver_diagrams}. Let ${\bf X}_{\rm OTFS}$ of size $M \times N$ be the information symbol matrix for OTFS, where $M$ denotes the number of delay bins/sub-carriers and $N$ is the number of Doppler bins/time slots. 
After passing through the IDZT module, the time domain OTFS symbol vector ${{\bf{s}}_{{\rm{OTFS}}}}$ of length $MN$ is given by 
\begin{align}
{{\bf{s}}_{{\rm{OTFS}}}} = \left( {{\bf{F}}_N^{\rm{H}} \otimes {{\bf{I}}_M}} \right){{\bf{x}}_{{\rm{OTFS}}}},
\label{OTFS_transmit_withoutCP}
\end{align}
where ${{\bf{x}}_{{\rm{OTFS}}}} = {\rm{vec}}\left( {{{\bf{X}}_{{\rm{OTFS}}}}} \right)$.
Adding the CP to ${{\bf{s}}_{{\rm{OTFS}}}}$, we obtain the transmitted OTFS symbol vector 
\begin{align}
{{{\bf{\tilde x}}}_{{\rm{OTFS}}}} = {\bf{A}}_{{\rm{CP}}}^{{\rm{OTFS}}} \left({{\bf{F}}_N^{\rm{H}} \otimes {{\bf{I}}_M}} \right) {{\bf{x}}_{{\rm{OTFS}}}},
\label{OTFS_transmit_withCP}
\end{align}
where ${\bf{A}}_{{\rm{CP}}}^{{\rm{OTFS}}} \in \mathbb{R}^{(MN + L_{\rm CP}^{\rm OTFS}) \times MN}$ is the CP addition matrix for OTFS. Let ${\bf{R}}_{{\rm{CP}}}^{{\rm{OTFS}}} \in \mathbb{R}^{MN \times (MN + L_{\rm CP}^{\rm OTFS})}$ be the CP removal matrix for OTFS. With $\bf w$ denoting AWGN as in~\eqref{SC_io_withoutCP_mtx}, the received time domain OTFS symbol vector ${{\bf{r}}_{{\rm{OTFS}}}}$ after removing the CP is given by
\begin{align}
{{\bf{r}}_{{\rm{OTFS}}}} = {\bf{R}}_{{\rm{CP}}}^{{\rm{OTFS}}}\sum\nolimits_{i = 1}^{{M_{{\rm{AP}}}}} {{{\bf{G}}^{\left( i \right)}}} {\bf{A}}_{{\rm{CP}}}^{{\rm{OTFS}}}{{\bf{s}}_{{\rm{OTFS}}}} + {\bf{w}}.
\label{OTFS_io_withoutCP_mtx_time}
\end{align}
Finally, by transforming the time domain vector ${{\bf{r}}_{{\rm{OTFS}}}}$ into the DD domain using the DZT, we arrive at the DD domain input-output relation
\begin{align}
{{\bf{y}}_{{\rm{OTFS}}}} =& \sum\nolimits_{i = 1}^{{M_{{\rm{AP}}}}} {\left({{\bf{F}}_N^{\rm{H}} \otimes {{\bf{I}}_M}}\right){\bf{R}}_{{\rm{CP}}}^{{\rm{OTFS}}}{{\bf{G}}^{\left( i \right)}}} \notag\\
& \times {\bf{A}}_{{\rm{CP}}}^{{\rm{OTFS}}}\left({{\bf{F}}_N^{\rm{H}} \otimes {{\bf{I}}_M}}\right){{\bf{x}}_{{\rm{OTFS}}}} + {\bf{w}}.
\label{OTFS_io_withoutCP_mtx}
\end{align} 

\section{Reduced-Complexity Detection for Different Waveforms}
In this section, we will present the considered reduced-complexity detection algorithms for three waveforms based on~\eqref{SC_io_withoutCP_mtx},~\eqref{OFDM_io_withoutCP_mtx}, and~\eqref{OTFS_io_withoutCP_mtx}, where the impact of the residual CFO as part of the effective Doppler is considered. Specifically, we consider FDE for both SC and OFDM, and a reduced-complexity CDID scheme for OTFS. All the above detection schemes are essentially single-tap FDEs with some variations. Note that the impact of Doppler becomes more significant with an increased frame length, which can be mitigated by adding CPs between adjacent blocks, such that \textit{low-complexity} block-based equalizers are sufficient. Since no such CP is added for OTFS, it has to rely on 
\textit{high-complexity} equalizers for improving communication reliability. In this section, we assume that the channel state information (CSI) is perfectly known at the receiver.

\subsection{Frequency-Domain Equalization for Single Carrier}
For simplicity, we consider the FDE for SC to be carried out in a block-by-block manner
after removing the CPs at the receiver side.
Specifically, for the $n$-th block, FDE is performed based on the frequency domain symbol vector ${\bf{q}}_{{\rm{SC}}}^{\left( n \right)}$ of length $L_{\rm SC}$ given by 
\begin{align}
{\bf{q}}_{{\rm{SC}}}^{\left( n \right)} \buildrel \Delta \over = {{\bf{F}}_{{L_{{\rm{SC}}}}}}{\bf{y}}_{{\rm{SC}}}^{\left( n \right)} = {{\bf{F}}_{{L_{{\rm{SC}}}}}}{\bf{G}}_{{\rm{SC}}}^{\left( n \right)}{\bf{x}}_{{\rm{SC}}}^{\left( n \right)} + {{\bf{w}}^{\left( n \right)}},
\label{SC_freq_mtx1}
\end{align}
where ${\bf{x}}_{{\rm{SC}}}^{\left( n \right)}$, ${\bf{y}}_{{\rm{SC}}}^{\left( n \right)}$, and ${{\bf{w}}^{\left( n \right)}}$ of length $L_{\rm SC}$ are the $n$-th block of ${\bf{x}}_{{\rm{SC}}}$, ${\bf{y}}_{{\rm{SC}}}$, and ${\bf{w}}$, respectively, and ${\bf{G}}_{{\rm{SC}}}^{\left( n \right)}$ of size $L_{\rm SC} \times L_{\rm SC}$ is the effective channel matrix for the $n$-th block{\footnote{Here, the same notation for the noise vector as in the time domain is used as they have the same distribution.}}. Specifically, ${\bf{ G}}_{{\rm{SC}}}^{\left( n \right)}$ is given by 
\begin{align}
{\bf{ G}}_{{\rm{SC}}}^{\left( n \right)} = {\bf{\tilde R}}_{{\rm{CP}}}^{{\rm{SC}}}{\bf{\tilde G}}_{{\rm{SC}}}^{\left( n \right)}{\bf{\tilde A}}_{{\rm{CP}}}^{{\rm{SC}}},
\label{SC_k_effective_channel_mtx}
\end{align}
where ${\bf{\tilde  G}}_{{\rm{SC}}}^{\left( n \right)}$ of size $(L_{\rm CP}+L_{\rm SC})\times (L_{\rm CP}+L_{\rm SC})$ is the $n$-th diagonal block of $\sum\nolimits_{i = 1}^{{M_{{\rm{AP}}}}} {{{\bf{G}}^{\left( i \right)}}} $.
In~\eqref{SC_k_effective_channel_mtx}, ${\bf{\tilde A}}_{{\rm{CP}}}^{{\rm{SC}}} \buildrel \Delta \over = {\left[ {{\bf{C}}_{{\rm{CP}}}^{{\rm{SC}}},{{\bf{I}}_{{L_{{\rm{SC}}}}}}} \right]^{\rm{T}}}$ is the block-wise CP addition matrix for the SC waveform, where ${{\bf{C}}_{{\rm{CP}}}^{{\rm{SC}}}}$ of size $L_{\rm SC} \times L_{\rm CP}$ contains the last $L_{\rm CP}$ columns of the identity matrix ${{\bf{I}}_{{L_{{\rm{SC}}}}}}$~\cite{RezazadehReyhani2018analysis}. Correspondingly, ${\bf{\tilde R}}_{{\rm{CP}}}^{{\rm{SC}}}$ is the block-wise CP reduction matrix for the SC waveform, which is of size $L_{\rm SC} \times (L_{\rm CP}+L_{\rm SC})$, obtained by removing the first $L_{\rm CP}$ rows of ${\bf I}_{L_{\rm CP}+L_{\rm SC}}$.  Let ${\bf{H}}_{{\rm{SC}}}^{\left( n \right)} = {{\bf{F}}_{{L_{{\rm{SC}}}}}}{\bf{ G}}_{{\rm{SC}}}^{\left( n \right)}{\bf{F}}_{{L_{{\rm{SC}}}}}^{\rm{H}}$ be the effective frequency domain channel matrix for SC. 
For notational brevity, we henceforth use $h_{i,j}^{\rm SC}$ denoting the $(i,j)$-th element in ${\bf{H}}_{{\rm{SC}}}^{\left( n \right)}$, where the superscript $(k)$ is dropped as the detection will be performed in a block-by-block manner. Let ${\bf z}_{\rm SC}^{\left(k\right)}={\bf{F}}_{{L_{{\rm{SC}}}}}{\bf x}_{\rm SC}^{(k)}$ be the equivalent frequency domain symbol vector. 
Then,~\eqref{SC_freq_mtx1} can be equivalently expressed by 
\begin{align}
{\bf{q}}_{{\rm{SC}}}^{\left( n \right)} = {\bf{H}}_{{\rm{SC}}}^{\left( n \right)}{{\bf{F}}_{{L_{{\rm{SC}}}}}}{\bf{x}}_{{\rm{SC}}}^{\left( n \right)} + {{\bf{w}}^{\left( n \right)}}={\bf{H}}_{{\rm{SC}}}^{\left( n \right)}{\bf{z}}_{{\rm{SC}}}^{\left( n \right)} + {{\bf{w}}^{\left( n \right)}},
\label{SC_freq_mtx2}
\end{align}
where the $l$-th symbol ${{q}}_{{\rm{SC}}}^{\left( n \right)}[l]$ of ${\bf{q}}_{{\rm{SC}}}^{\left( n \right)}$ satisfies 
\begin{align}
q_{{\rm{SC}}}^{\left( n \right)}\left[ l \right]\! =\! h_{l,l}^{{\rm{SC}}}z_{{\rm{SC}}}^{\left( n \right)}\left[ l \right] \!+\! \sum\nolimits_{\scriptstyle j = 0\hfill\atop
\scriptstyle j \ne l\hfill}^{MN - 1} {h_{l,j}^{{\rm{SC}}}z_{{\rm{SC}}}^{\left( n \right)}\left[ j \right]}  \!+ \!{w^{\left( n \right)}}\left[ l \right].
\label{SC_freq_symbol_l}
\end{align}
Based on~\eqref{SC_freq_symbol_l}, we apply the single-tap equalization according to the MMSE criterion, which can be characterized by a diagonal matrix ${\bf{W}}_{{\rm{SC}}}^{\left( n \right)}$. Specifically, the $l$-th MMSE filter coefficient, i.e., the $l$-th diagonal entry of ${\bf{W}}_{{\rm{SC}}}^{\left( n \right)}$, is given by
\begin{align}
W_{{\rm{SC}}}^{\left( n \right)}\left[ {l,l} \right] = \frac{{{E_s}{{\left( {h_{l,l}^{{\rm{SC}}}} \right)}^*}}}{{ {E_s}\sum\nolimits_{\scriptstyle j = 0}^{MN - 1} {{{\left| {h_{l,j}^{{\rm{SC}}}} \right|}^2}}  + {N_0}}},
\label{SC_MMSE_l}
\end{align}
where $E_s$ is the average symbol energy. 
Thus, the estimated frequency domain symbol vector ${\bf{\hat q}}_{{\rm{SC}}}^{\left( n \right)}$ is given by ${\bf{\hat q}}_{{\rm{SC}}}^{\left( n \right)}={\bf{W}}_{{\rm{SC}}}^{\left( n \right)}{\bf{q}}_{{\rm{SC}}}^{\left( n \right)}$.
Finally, the estimated information symbol vector ${\bf{\hat x}}_{{\rm{SC}}}^{\left( n \right)}$ for the $n$-th block can be obtained by transforming~${\bf{\hat q}}_{{\rm{SC}}}^{\left( n \right)}$ back into the time domain, i.e., ${\bf{\hat x}}_{{\rm{SC}}}^{\left( n \right)} ={\bf{F}}_{{L_{{\rm{SC}}}}}^{\rm{H}}{\bf{\hat q}}_{{\rm{SC}}}^{\left( n \right)}$.

\subsection{Frequency-Domain Equalization for OFDM}
For OFDM, the single-tap FDE is the \emph{de facto} choice for receiver processing. Similar to the previous subsection, the considered FDE is performed for each OFDM symbol individually. Let ${\bf x}_{\rm OFDM}^{\left( n \right)}$ and ${\bf y}_{\rm OFDM}^{\left( n \right)}$ of length $M_{\rm OFDM}$ be the $n$-th block of ${\bf{x}}_{{\rm{OFDM}}}$ and ${\bf{y}}_{{\rm{OFDM}}}$, respectively. Then, according to~\eqref{OFDM_io_withoutCP_mtx}, we have 
\begin{align}
{\bf y}_{\rm OFDM}^{\left( n \right)}={\bf{H}}_{{\rm{OFDM}}}^{\left( n \right)} {\bf x}_{\rm OFDM}^{\left( n \right)}+{{\bf{w}}^{\left( n \right)}},
\label{OFDM_equ_mtx}
\end{align}
where ${\bf{H}}_{{\rm{OFDM}}}^{\left( n \right)}$ of size $M_{\rm OFDM} \times M_{\rm OFDM}$ is the effective frequency domain channel matrix for OFDM. Similar to the previous subsection, ${\bf{H}}_{{\rm{OFDM}}}^{\left( n \right)}$ can be obtained by the multiplication of DFT matrices, i.e., ${\bf{H}}_{{\rm{OFDM}}}^{\left( n \right)} = {{\bf{F}}_{{M_{{\rm{OFDM}}}}}}{\bf{\tilde G}}_{{\rm{OFDM}}}^{\left( n \right)}{\bf{F}}_{{M_{{\rm{OFDM}}}}}^{\rm{H}}$, where ${\bf{\tilde G}}_{{\rm{OFDM}}}^{\left( n \right)} \in \mathbb{C}^{M_{\rm OFDM} \times M_{\rm OFDM}}$ is the effective time domain channel matrix for OFDM after removing the CP. Specifically, ${\bf{\tilde G}}_{{\rm{OFDM}}}^{\left( n \right)}$ is obtained as in~\eqref{SC_k_effective_channel_mtx} by replacing the matrix dimensions with the corresponding OFDM parameters. 
Similar to SC,
we henceforth use $h_{i,j}^{\rm OFDM}$ denoting the $(i,j)$-th element in ${\bf{H}}_{{\rm{OFDM}}}^{\left( n \right)}$. Then, the $l$-th symbol of ${\bf y}_{\rm OFDM}^{\left( n \right)}$ satisfies
\begin{align}
y_{{\rm{OFDM}}}^{\left( n \right)}\left[ l \right] &= h_{l,l}^{{\rm{OFDM}}}x_{{\rm{OFDM}}}^{\left( n \right)}\left[ l \right] \nonumber \\ 
&\quad + \sum\nolimits_{\scriptstyle j = 0\hfill\atop
\scriptstyle j \ne l\hfill}^{MN - 1} {h_{l,j}^{{\rm{OFDM}}}x_{{\rm{OFDM}}}^{\left( n \right)}\left[ j \right]}  + {w^{\left( n \right)}}\left[ l \right].
\label{OFDM_freq_symbol_l}
\end{align}
Let us then consider the single-tap equalization according to the MMSE criterion for OFDM using the diagonal matrix ${\bf{W}}_{{\rm{OFDM}}}^{\left( n \right)}$, whose $l$-th diagonal entry is given by
{\color{black}$W_{{\rm{OFDM}}}^{\left( n \right)}\left[ {l,l} \right] = \frac{{{E_s}{{\left( {h_{l,l}^{{\rm{OFDM}}}} \right)}^*}}}{{{E_s}\sum\nolimits_{\scriptstyle j = 0}^{MN - 1} {{{\left| {h_{l,j}^{{\rm{OFDM}}}} \right|}^2}}  + {N_0}}}$.}
Finally, by applying ${\bf{W}}_{{\rm{OFDM}}}^{\left( n \right)}$ to ${\bf y}_{{\rm{OFDM}}}^{\left( n \right)}$, we obtain the estimated information symbol vector ${\bf{\hat x}}_{{\rm{OFDM}}}^{\left( n \right)}$ for the $n$-th block by ${\bf{\hat x}}_{{\rm{OFDM}}}^{\left( n \right)}={\bf W}_{{\rm{OFDM}}}^{\left( n \right)}{\bf y}_{{\rm{OFDM}}}^{\left( n \right)}$.

\subsection{FDE-based CDID for OTFS}
\begin{figure}
\centering
\includegraphics[width=.99\linewidth]{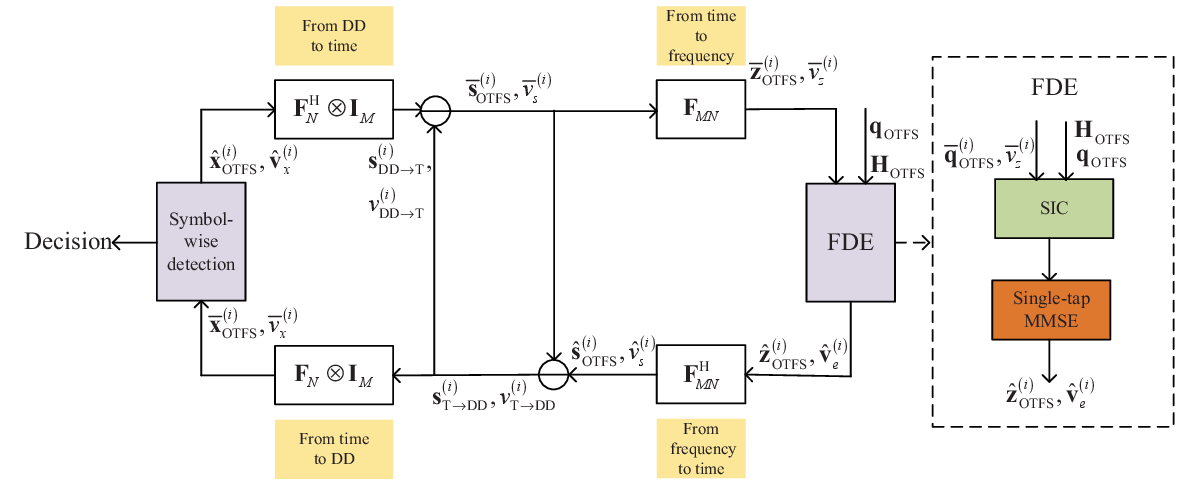}
\caption{The block diagram for the considered FDE-based CDID for OTFS.}
\label{cross_domain}
\centering
\vspace{-3mm}
\end{figure}
Different from the existing CDID methods~\cite{li2021cross,Mengmeng2023crossdomain} that applies the time-domain MMSE, the proposed scheme particularly considers FDE motivated by the small effective Doppler. The diagram of the proposed CDID scheme is given in Fig. \ref{cross_domain}. The proposed scheme consists of an estimator of time domain symbols and a detector of DD domain symbols that are connected via a domain transformation and an extrinsic information calculation.  
Let us define the effective time domain channel matrix for OTFS
${\bf G}_{\rm OTFS}= {\bf R}_{\rm CP}^{\rm OTFS}\sum\nolimits_{i = 1}^{{M_{{\rm{AP}}}}} {{{\bf{G}}^{\left( i \right)}}}{\bf A}_{\rm CP}^{\rm OTFS}$,
where ${\bf R}_{\rm CP}^{\rm OTFS}$ and ${\bf A}_{\rm CP}^{\rm OTFS}$ can be obtained like ${\bf{\tilde A}}_{{\rm{CP}}}^{{\rm{SC}}}$ and ${\bf{\tilde R}}_{{\rm{CP}}}^{{\rm{SC}}}$ by using the corresponding OTFS parameters.
Based on ${\bf G}_{\rm OTFS}$, let us define the frequency domain channel matrix for OTFS by ${\bf{H}}_{{\rm{OTFS}}} \triangleq {{\bf{F}}_{MN}}{\bf{G}}_{{\rm{OTFS}}}{\bf{F}}_{MN}^{\rm{H}}$, whose $(i,j)$-th element is denoted by $h_{i,j}^{\rm OTFS}$. Furthermore, let ${\bf z}={\bf{F}}_{MN}{\bf s}_{\rm OTFS}$ be the equivalent frequency domain symbol vector and we have the received frequency domain symbol vector for OTFS written as
${\bf{q}}_{{\rm{OTFS}}} = {{\bf{F}}_{{MN}}}{\bf{r}}_{{\rm{OTFS}}} = {\bf{H}}_{{\rm{OTFS}}}{\bf{z}}+ {\bf{w}}$,
where the $l$-th entry
of ${\bf{q}}_{{\rm{OTFS}}}$ satisfies 
\begin{align}
{q_{{\rm{OTFS}}}}\left[ l \right]\! = \!h_{l,l}^{{\rm{OTFS}}}{z}\left[ l \right] \!+\! \sum\nolimits_{\scriptstyle j = 0\hfill\atop
\scriptstyle j \ne l\hfill}^{MN - 1} {h_{l,j}^{{\rm{OTFS}}}{z}\left[ j \right]} \! +\! w\left[ l \right].
\label{OTFS_freq_symbol_l}
\end{align}

The proposed FDE-based CDID is derived from~\eqref{OTFS_freq_symbol_l} by considering successive interference cancellation (SIC) in the frequency domain based on the previous iterations{\footnote{Note that a similar CDID scheme may be applied for the SC, which will be considered in future work. In this paper, we only consider standard FDE for SC with inserted CP for improving the detection performance. }}. For simplicity, we assume that the time domain symbols are Gaussian distributed with different means but same variance, which is reasonable due to the dense unitary matrices characterizing the domain transformation~\cite{Ruoxi_CDID_ITW}. 
Specifically, in the $i$-th iteration, the proposed algorithm starts in the time domain with the \textit{a priori} mean of the time domain symbol vector ${\bar {\bf s}}^{(i)}_{\rm OTFS}$ and corresponding variance ${\bar v}_s^{(i)}$ as follows. 

\subsubsection{From time domain to frequency domain} Derive the \textit{a priori} mean and variance of frequency domain symbols\footnote{In the first iteration, we assume that the time domain symbol is zero mean with variance $E_s$.}, i.e., ${{\bf\bar z}_{{\rm{OTFS}}}^{(i)}} = {{\bf{F}}_{MN}}{{{\bf{\bar s}}}_{{\rm{OTFS}}}^{(i)}}$ and ${\bar v}_z^{(i)}={\bar v}_s^{(i)}$.

\subsubsection{FDE with SIC} Calculate the MMSE coefficient for the $l$-th frequency symbol by
    \begin{align}
    {W_{{\rm{OTFS}}}}\left[ {l,l} \right] = \frac{{{{{{\bar v}_z^{(i)}}}}{{\left( {h_{l,l}^{{\rm{OTFS}}}} \right)}^*}}}{{ {{\bar v}_z^{(i)}}\sum\nolimits_{\scriptstyle j = 0}^{MN - 1} {{{\left| {h_{l,j}^{{\rm{OTFS}}}} \right|}^2}}  + {N_0}}}
    \label{OTFS_MMSE_l}.
    \end{align}
    Thus, the \textit{a posteriori} mean is given by
    \begin{align}
    {{\hat z}_{{\rm{OTFS}}}^{(i)}}\!\left[ l \right]\! = \!\!{{\bar z}_{{\rm{OTFS}}}^{(i)}}\!\left[ l \right]\! \!+\! \!{W_{{\rm{OTFS}}}}\!\left[ {l\!,\!l} \right]\!\left( {{q_{{\rm{OTFS}}}}\!\left[ l \right] \!-\! {{\bar q}_{{\rm{OTFS}}}^{(i)}}\!\left[ l \right]} \right),
    \label{OTFS_freq_MMSE_mean}
    \end{align}
    where ${{\bar q}_{{\rm{OTFS}}}^{(i)}}\left[ l \right]$ is the $l$-th entry of ${{\bf\bar q}_{{\rm{OTFS}}}^{(i)}} = {{\bf{H}}_{{\rm{OTFS}}}}{{{\bf{\bar z}}}_{{\rm{OTFS}}}^{(i)}}$. The corresponding error variance is obtained as
    \begin{align}
    {{\hat v}_e}\left[ l \right] = {{\bar v}_z^{(i)}}- {{\bar v}_z^{(i)}}\frac{{{{\bar v}_z^{(i)}}{{\left| {h_{l,l}^{{\rm{OTFS}}}} \right|}^2}}}{{{{\bar v}_z^{(i)}}\sum\nolimits_{j = 0}^{MN - 1} {{{\left| {h_{l,j}^{{\rm{OTFS}}}} \right|}^2} + {N_0}} }}.
    \label{OTFS_freq_MMSE_var}
    \end{align}
\subsubsection{From frequency domain to time domain} Derive the \textit{a posteriori} mean and variance of time domain symbols, i.e., ${{\bf\hat s}_{{\rm{OTFS}}}^{(i)}} = {\bf F}_{MN}^{\rm H}{{{\bf{\hat z}}}_{{\rm{OTFS}}}^{(i)}}$ and ${\hat v}_s^{(i)}={\mathbb E}\left[{\bf \hat v}_e^{(i)}\right]$.    

\subsubsection{Cross domain message passing (time to DD)} Calculate the \textit{extrinsic} information in the time domain~\cite{li2021cross}, i.e., ${v}^{(i)}_{s, \rm T \rightarrow DD}= \left( \frac{1}{\hat{v}^{(i)}_s} - \frac{1}{\bar{v}^{(i)}_s} \right)^{-1}$ and ${ s}^{(i)}_{\rm T \rightarrow DD}[l] = {v}^{(i)}_{s, \rm T \rightarrow DD} \left( \frac{ \hat{s}^{(i)}[l] }{ \hat{v}^{(i)}_s } - \frac{ \bar{s}^{(i)}[l] }{ \bar{v}^{(i)}_s }     \right) $. Then, the extrinsic information is passed to the DD domain obtaining the \textit{a priori} information for the DD domain symbols, i.e., ${{{\bf{\bar x}}}^{(i)}_{{\rm{OTFS}}}} = \left( {{{\bf{F}}_N} \otimes {{\bf{I}}_M}} \right){\bf s}^{(i)}_{\rm T \rightarrow DD}$
and ${\bar v}_x^{(i)}={v}^{(i)}_{s, \rm T \rightarrow DD}$.

\subsubsection{DD domain detection} The DD domain detection can be straightforwardly implemented by calculating the \textit{a posteriori} probability (APP) based on the Euclidean distances between the elements in ${{{\bf{\bar x}}}_{{\rm{OTFS}}}^{(i)}}$ and the possible constellation points~\cite{li2021cross}. The DD domain detector returns the mean and variance of each DD domain symbol, denoted as ${\hat x}_{\rm OTFS}^{(i)}\left[l\right]$ and ${{{\hat v}}_x^{(i)}}\left[l\right]$.

\subsubsection{Cross domain message passing (DD to time)} Pass the \textit{a posteriori} information from the DD domain to the time domain by ${ {\bf s}}^{(i)}_{\rm DD \rightarrow T}=\left( {{{\bf{F}}_N^{\rm H}} \otimes {{\bf{I}}_M}} \right){\bf \hat x}_{\rm OTFS}$ and corresponding variance $v_{s, \rm DD \rightarrow T}^{(i)}={\mathbb E}\left[{\bf \hat v}_x^{(i)}\right]$. Then, calculate the \textit{a priori} information of time domain symbols for the forthcoming iteration in the form of extrinsic information, i.e., ${\bar v}^{(i+1)}_{s}= \left( \frac{1}{v_{s, \rm DD \rightarrow T}^{(i)}} - \frac{1}{v_{s, \rm T \rightarrow DD}^{(i)}} \right)^{-1}$ and ${\bar s}^{(i+1)}_{\rm OTFS}[l] = {\bar v}^{(i+1)}_{s} \left( \frac{ { { s}}^{(i)}_{\rm DD \rightarrow T}[l] }{ {v}^{(i)}_{s, \rm DD \rightarrow T} } - \frac{ { { s}}^{(i)}_{\rm T \rightarrow DD}[l] }{ {v}^{(i)}_{s, \rm T \rightarrow DD} }    \right) $.

\textbf{Remark on complexity:} The proposed FDE-based CDID clearly demands higher complexity compared to the FDE counterparts for SC and OFDM waveforms due to the iterative nature and the larger frame size. Specifically, the FDE for SC requires two fast Fourier transform (FFT) operations per block having overall complexity in the order of ${\cal O}\left(2NL_{\rm SC}\log(L_{\rm SC})\right)$. Furthermore, the FDE for OFDM only requires one FFT operation per OFDM symbol and therefore its complexity is in the order of ${\cal O}\left(NM_{\rm OFDM}\log(M_{\rm OFDM})\right)$. In contrast, within each iteration, the proposed CDID requires complexity in the order of 
${\cal O}\left(2MN\log(MN)+2MN\log(N)\right)$.

\section{Numerical Results}
Without loss of generality, we consider $M_{\rm AP} =4$, $M=32$, and $N=16$ for the three waveforms, where QPSK signaling is applied. We assume that the maximum delay and Doppler after pre-compensation at the AP correspond to at most $9.3$\% and $1.5$\% of the symbol duration and subcarrier spacing, yielding the same
CP length $L_{\rm CP}^{\rm SC}=L_{\rm CP}^{\rm OFDM}=L_{\rm CP}^{\rm OTFS}=3$ for all waveforms. To evaluate the performance, we consider the ``pragmatic capacity", which is defined as a single-letter mutual information incorporating the effect of the modulation, channel, and equalization~\cite{Fair_compare_OTFS_OFDM}. The simulations are done for a sufficient number of transmissions with different \textit{transmit signal-to-noise ratio (SNR)} defined here as the ratio between the average energy per symbol per AP and the complex baseband noise power spectral density $N_0$.

\begin{figure}[tp]
\centering
\includegraphics[width=0.8\linewidth]{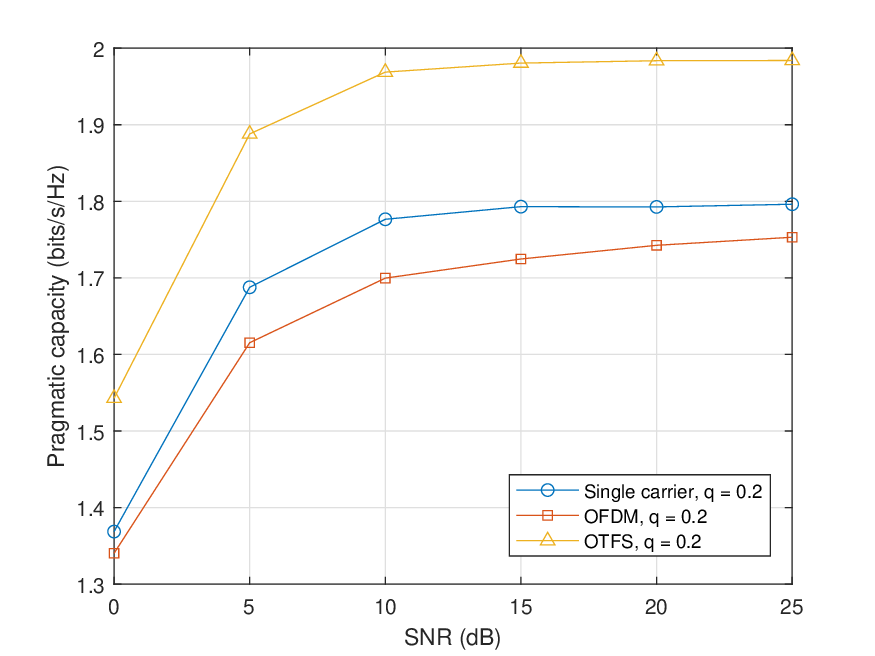}
\caption{The comparison among three waveforms in terms of the pragmatic capacity, where the blocking rate is $q=0.2$.}
\label{PC_figure}
\centering
\vspace{-1mm}
\end{figure}
We present the comparison among three waveforms in terms of the pragmatic capacity \cite{Fair_compare_OTFS_OFDM} in Fig.~\ref{PC_figure}, where the blocking rate is $q=0.2$ and the results are averaged over multiple transmissions. We observe that OTFS exhibits the best performance among the three waveforms, while OFDM is the worst. Specifically, the significant performance improvement of OTFS results from a better detection performance and less CP overhead compared to the other two waveforms. Furthermore, SC slightly outperforms OFDM. This is because the information symbols of the SC waveform are placed in the time domain, and the residual frequency domain interference after FDE is potentially averaged out in the time domain due to the dense DFT matrix.

\begin{figure}[tp]
\centering
\includegraphics[width=0.8\linewidth]{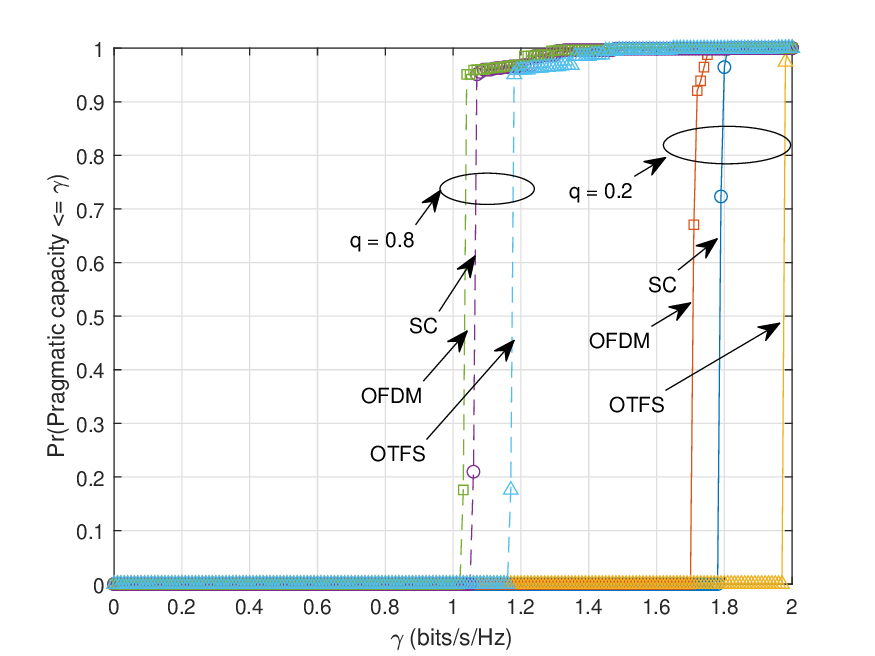}
\caption{Empirical CDF of the pragmatic capacity for multiple transmissions with $q=0.2$ and $q=0.8$.}
\label{CDF_figure}
\centering
\vspace{-1mm}
\end{figure}

The cumulative distribution functions (CDFs) of the pragmatic capacities for the three waveforms are demonstrated in Fig.~\ref{CDF_figure}, where blocking rates of $q=0.2$ and $q=0.8$ are considered and the underlying transmit SNR is $10$ dB. As shown in the figure, all the waveforms exhibit better pragmatic capacities with a lower blocking rate. Furthermore, OTFS outperforms the other two waveforms for both considered blocking rates, which align well with the results from Fig.~\ref{PC_figure}.

\section{Conclusions}
This paper studied the performance of typical waveforms for mmWave downlink transmissions with multi-connectivity. A reduced-complexity detector based on the FDE was proposed for OTFS following the CDID framework. Our numerical results verified the effectiveness of the proposed OTFS detector and demonstrated better pragmatic capacity performance at the cost of increased detection complexity. 

\section{Acknowledgments}
F. G\"ottsch, L. Miretti, G. Caire and S. Sta\'nczak acknowledge the financial support by the Federal Ministry of Education and Research of Germany in the programme of “Souverän. Digital. Vernetzt.” Joint project 6G-RIC, project identification number: 16KISK020K (L. Miretti and S. Sta\'nczak) and 16KISK030 (all mentioned authors).

The work of S. Li is supported in part by the European Union’s Horizon 2020 Research and Innovation Program under MSCA Grant No. 101105732 – DDComRad.

\bibliographystyle{IEEEtran}
\bibliography{OTFS_references}

\begin{thebibliography}{10}
\providecommand{\url}[1]{#1}
\csname url@samestyle\endcsname
\providecommand{\newblock}{\relax}
\providecommand{\bibinfo}[2]{#2}
\providecommand{\BIBentrySTDinterwordspacing}{\spaceskip=0pt\relax}
\providecommand{\BIBentryALTinterwordstretchfactor}{4}
\providecommand{\BIBentryALTinterwordspacing}{\spaceskip=\fontdimen2\font plus
\BIBentryALTinterwordstretchfactor\fontdimen3\font minus
  \fontdimen4\font\relax}
\providecommand{\BIBforeignlanguage}[2]{{%
\expandafter\ifx\csname l@#1\endcsname\relax
\typeout{** WARNING: IEEEtran.bst: No hyphenation pattern has been}%
\typeout{** loaded for the language `#1'. Using the pattern for}%
\typeout{** the default language instead.}%
\else
\language=\csname l@#1\endcsname
\fi
#2}}
\providecommand{\BIBdecl}{\relax}
\BIBdecl

\bibitem{ngo2024ultradense}
H.~Q. Ngo, G.~Interdonato, E.~G. Larsson, G.~Caire, and J.~G. Andrews,
  ``{Ultradense Cell-Free Massive MIMO for 6G: Technical Overview and Open
  Questions},'' \emph{Proc. IEEE}, 2024.

\bibitem{Caire-Shamai-TIT03}
G.~Caire and S.~{Shamai (Shitz)}, ``{On the achievable throughput of a
  multiantenna Gaussian broadcast channel},'' \emph{IEEE Trans. Inf. Theory},
  vol.~49, no.~7, pp. 1691--1706, Jul. 2003.

\bibitem{Marzetta-TWC10}
T.~L. Marzetta, ``{Noncooperative cellular wireless with unlimited numbers of
  base station antennas},'' \emph{{IEEE Trans. on Wireless Commun.}}, vol.~9,
  no.~11, pp. 3590--3600, Nov. 2010.

\bibitem{miretti2024robust}
L.~Miretti, G.~Caire, and S.~Stańczak, ``{Robust mmWave/sub-THz
  Multi-Connectivity Using Minimal Coordination and Coarse Synchronization},''
  \emph{{IEEE Trans. on Wireless Commun.}}, vol.~24, no.~1, pp. 293--308, 2025.

\bibitem{irmer2011coordinated}
R.~Irmer, H.~Droste, P.~Marsch, M.~Grieger, G.~Fettweis, S.~Brueck, H.-P.
  Mayer, L.~Thiele, and V.~Jungnickel, ``{Coordinated multipoint: Concepts,
  performance, and field trial results},'' \emph{IEEE Commun. Mag.}, vol.~49,
  no.~2, pp. 102--111, Feb. 2011.

\bibitem{you2020network}
L.~You, X.~Chen, X.~Song, F.~Jiang, W.~Wang, X.~Gao, and G.~Fettweis,
  ``{Network massive MIMO transmission over millimeter-wave and terahertz
  bands: Mobility enhancement and blockage mitigation},'' \emph{IEEE Jour. Sel.
  Areas Commun.}, vol.~38, no.~12, pp. 2946--2960, Dec. 2020.

\bibitem{Li2020performance}
S.~Li, J.~Yuan, Z.~Wei, B.~Bai, and D.~W.~K. Ng, ``Performance analysis of
  coded {OTFS} systems over high-mobility channels,'' \emph{IEEE Trans.
  Wireless Commun.}, vol.~20, no.~9, pp. 6033--6048, Sep. 2021.

\bibitem{mohammadi2022cell}
M.~Mohammadi, H.~Q. Ngo, and M.~Matthaiou, ``{Cell-free massive MIMO meets OTFS
  modulation},'' \emph{IEEE Trans. Commun.}, vol.~70, no.~11, pp. 7728--7747,
  2022.

\bibitem{buzzi2023leo}
S.~Buzzi, G.~Caire, G.~Colavolpe, C.~D’Andrea, T.~Foggi, A.~Piemontese, and
  A.~Ugolini, ``{LEO Satellite Diversity in 6G Non-Terrestrial Networks: OFDM
  vs. OTFS},'' \emph{IEEE Commun. Lett.}, vol.~27, no.~11, pp. 3013--3017,
  2023.

\bibitem{miretti2024little}
L.~Miretti, T.~K{\"u}hne, A.~Schultze, W.~Keusgen, G.~Caire, M.~Peter,
  S.~Sta{\'n}czak, and T.~Eichler, ``{Little or no equalization is needed in
  energy-efficient sub-THz mobile access},'' \emph{IEEE Commun. Mag.}, vol.~62,
  no.~2, pp. 94--100, Feb. 2024.

\bibitem{li2021cross}
S.~Li, W.~Yuan, Z.~Wei, and J.~Yuan, ``Cross domain iterative detection for
  orthogonal time frequency space modulation,'' \emph{IEEE Trans. Wireless
  Commun.}, vol.~21, no.~4, pp. 2227--2242, Apr. 2022.

\bibitem{Mengmeng2023crossdomain}
M.~Liu, S.~Li, B.~Bai, and G.~Caire, ``Reduced-complexity cross-domain
  iterative detection for {OTFS} modulation via delay-{Doppler} decoupling,''
  in \emph{Proc. 2023 IEEE Int. Workshop Signal Process. Adv. Wireless
  Commun.(SPAWC)}, 2023, pp. 1--6.

\bibitem{shokri2015millimeter}
H.~Shokri-Ghadikolaei, C.~Fischione, G.~Fodor, P.~Popovski, and M.~Zorzi,
  ``{Millimeter wave cellular networks: A MAC layer perspective},'' \emph{IEEE
  Trans. Commun.}, vol.~63, no.~10, pp. 3437--3458, Oct. 2015.

\bibitem{Raviteja2019practical}
P.~{Raviteja}, Y.~{Hong}, E.~{Viterbo}, and E.~{Biglieri}, ``Practical
  pulse-shaping waveforms for reduced-cyclic-prefix {OTFS},'' \emph{IEEE Trans.
  Veh. Technol.}, vol.~68, no.~1, pp. 957--961, Jan. 2019.

\bibitem{lampel2021orthogonal}
F.~Lampel, H.~Joudeh, A.~Alvarado, and F.~M.~J. Willems, ``Orthogonal time
  frequency space modulation based on the discrete {Zak} transform,''
  \emph{Entropy}, vol.~24, no.~12, Nov. 2022.

\bibitem{RezazadehReyhani2018analysis}
A.~{RezazadehReyhani}, A.~{Farhang}, M.~{Ji}, R.~R. {Chen}, and
  B.~{Farhang-Boroujeny}, ``Analysis of discrete-time {MIMO} {OFDM}-based
  orthogonal time frequency space modulation,'' in \emph{Proc. 2018 IEEE Int.
  Conf. Commun.}, May 2018, pp. 1--6.

\bibitem{Ruoxi_CDID_ITW}
R.~Chong, S.~Li, Z.~Wei, M.~Matthaiou, D.~W.~K. Ng, and G.~Caire, ``Analysis of
  cross-domain message passing for {OTFS} transmissions,'' in \emph{IEEE Inf.
  Theory Workshop}, 2024, pp. 1--6.

\bibitem{Fair_compare_OTFS_OFDM}
L.~Gaudio, G.~Colavolpe, and G.~Caire, ``{OTFS vs. OFDM} in the presence of
  sparsity: A fair comparison,'' \emph{IEEE Trans. Wireless Commun.}, vol.~21,
  no.~6, pp. 4410--4423, Dec. 2022.

\end{thebibliography}

\end{document}